\documentclass[%
 reprint,
 superscriptaddress,
 amsmath,amssymb,
 aps,
 prb,
floatfix,
longbibliography
]{revtex4-2}

\usepackage{graphicx}%
\usepackage{dcolumn}%
\usepackage{bm}%
\usepackage{color}%
\usepackage{siunitx}
\usepackage{here}
\usepackage{mathtools}
\usepackage{hyperref}

\begin{document}
\preprint{APS/123-QED}

\title{Universal Magnetoresistance Scaling in Layered Pd-based Multiband Metals Beyond Compensated Semimetal Regime}%

\author{K. Okawa}
\email{okawa.k@aist.go.jp}
\affiliation{%
Materials and Structures Laboratory, Institute of Science Tokyo, Yokohama 226-8501, Japan}%
\affiliation{%
National Metrology Institute of Japan (NMIJ), National Institute of Advanced Industrial Science and Technology (AIST), AIST Central 3, 1-1-1 Umezono, Tsukuba 305-8563, Japan}%
\author{T. Sasagawa}
\affiliation{%
Materials and Structures Laboratory, Institute of Science Tokyo, Yokohama 226-8501, Japan}%
\
\date{\today}

\begin{abstract}
We systematically investigated the magnetoresistance ratio ($MR$) of single-crystalline, nonmagnetic layered Pd-based metals, including centrosymmetric PdTe$_{2}$, PdPb$_{2}$, and $\beta$-PdBi$_{2}$ and noncentrosymmetric $\alpha$-PdBi. Our study identifies a distinct class of large $MR$ in multiband, high-carrier-density systems. Unlike well-studied extremely large $MR$ materials, such as Dirac and Weyl semimetals described by simple compensated-carrier models, these compounds possess complex Fermi surfaces, as validated by our first-principles calculations. Nevertheless, they exhibit a remarkably simple $MR$ scaling governed by carrier mobility, manifested in systematic dependencies on magnetic-field, temperature, and the residual resistivity ratio ($RRR$). The validity of Kohler’s rule in high-$RRR$crystals indicates that $MR$ is governed by a single effective scattering time, even in these multiband systems. The field and $RRR$dependences of $MR$ follow an intermediate power-law behavior between linear and quadratic, attributable to imperfect carrier compensation and a distribution of carrier mobilities. Among the studied compounds, $\alpha$-PdBi exhibits the largest $MR$, reaching $\SI{1.5e3}{\%}$ ($\SI{2}{K}$, $\SI{7}{T}$), owing to its exceptionally high $RRR$($\sim$660). However, when compared on an equal-$RRR$basis, its $MR$ is smaller than that of its centrosymmetric counterparts. This trend suggests that additional scattering channels, arising from spin-orbit-induced band splitting in noncentrosymmetric systems, reduce the effective carrier mobility. Our results establish a new class of large $MR$ in clean multiband metals where complex electronic structures give rise to emergent single-parameter scaling, highlighting the interplay between disorder, mobility, and symmetry.
\end{abstract}

\maketitle


\section{\label{sec:level1}Introduction}
Magnetoresistance, the field-induced change in electrical resistivity, is commonly quantified by the magnetoresistance ratio $MR = (\rho(H)-\rho(0))/\rho(0)$, which provides a sensitive probe of carrier dynamics and Fermi-surface topology in materials\cite{Pippard1989,Niu2021,Hu2008,Parish2003,Abrikosov1998}. Extremely large magnetoresistance (XMR), exceeding $\SI{e5}{\percent}$, has been observed in nonmagnetic compensated semimetals such as WTe$_{2}$, where near-perfect electron-hole compensation play a central role\cite{Ali2014,Ali2015,Shekhar2015,Leahy2018,Okazaki2026}. Within the semiclassical two-band framework, such systems typically exhibit a quadratic dependence of $MR$ on both magnetic field and carrier mobility, providing a benchmark for understanding X$MR$ phenomena. Importantly, however, large magnetoresistance is not limited to compensated semimetals. Even in carrier-rich metals, significant magnetoresistance can emerge when high carrier mobility is realized, and a complex Fermi-surface topology does not preclude such behavior\cite{Niu2021}. Because the mobility depends on both the intrinsic electronic structure (such as the effective mass) and extrinsic scattering, understanding which factor primarily controls the magnetotransport remains a key challenge.

A similar trend can be identified in superconducting materials. Superconductors with semimetallic electronic structures, such as MoTe$_{2}$ \cite{Chen2016,Rhodes2017}, $\gamma$-PtBi$_{2}$ \cite{Gao2018,Wu2020,Xing2020}, and CaSb$_{2}$ \cite{Funada2019,Oudau2020}, exhibit X$MR$ in their normal state, consistent with their proximity to carrier compensation. On the other hand, several well-established classes of multiband superconductors, including Fe-based superconductors (FeSe \cite{Watson2015,Wang2025}, FeS \cite{Lin2016}, FeTe$_{1-x}$Se$_{x}$ \cite{Sun2014}), BiS$_{2}$-based superconductors (Bi$_{4}$O$_{4}$S$_{3}$ \cite{Li2013} and CeO$_{1-x}$F$_{x}$BiS$_{2}$ \cite{Xing2012}), and noncentrosymmetric systems (Mg$_{12-\delta}$Ir$_{19}$B$_{16}$ \cite{Mu2010}, Nb$_{0.18}$Re$_{0.82}$ \cite{Sundar2019}, 4$H_{\rm b}$-TaS$_{2}$\cite{Gao2020}), typically show $MR$ that remains below 100\%. These materials are characterized by relatively large carrier densities and incomplete compensation, despite their multiband nature. These observations suggest that large magnetoresistance is not a generic consequence of multiband electronic structure, but instead reflects the underlying transport regime governed by carrier mobility and scattering. 

In addition to its magnitude, the scaling behavior and magnetic-field dependence of $MR$ provide further insight into the transport mechanism. While ideal compensated systems exhibit a quadratic field dependence, deviations toward linear or sub-quadratic behavior reflect departures from perfect compensation. When multiple electron and hole pockets coexist, the transport is governed by distinct carrier mobilities and scattering times across different bands, which is generally expected to cause deviations from simple scaling laws like Kohler's rule. Nevertheless, whether an emergent, simple magnetotransport description can arise in carrier-rich multiband metals remains largely unexplored. 

Another unresolved issue concerns the relationship between magnetoresistance and sample quality. Since carrier mobility is closely related to the residual resistivity ratio ($RRR$), systematic investigations of $MR-RRR$ scaling can provide direct insight into how disorder and scattering govern magnetotransport \cite{Ali2015,Okazaki2026}. Importantly, in the clean limit where defect scattering is very small, the $RRR$is no longer just a measure of impurity concentration; its absolute value can change significantly depending on intrinsic electronic parameters like the carrier effective mass. However, such studies focusing on the connection between $MR$, $RRR$, and the electronic structure remain scarce for multiband metals with complex Fermi surfaces.

Pd-based binary superconductors provide an ideal platform to explore these issues, as they host a variety of crystal structures with different symmetry properties and spin-orbit coupling (SOC) strength. Among them, $\alpha$-PdBi is a noncentrosymmetric monoclinic compound, enabling strong antisymmetric SOC and Rashba-type spin splitting \cite{Joshi2011,Okawa2013,Matano2013,Neupane2016,Benia2016,Thirupathaiah2016}. First-principles calculations indicate highly anisotropic, SOC-split multiband Fermi surfaces, suggesting unconventional magnetotransport behavior \cite{Lohani2017,Pramanik2021,Jiao2014,Sun2015,Peets2016,Yim2018,Cameron2025,Mondal2013,Klotz2020,Khan2019,Pramanik2020,Yaresko2018}. In contrast, $\beta$-PdBi$_{2}$\cite{Imai2012,Sakano2015,Iwaya2017,Powell2025,Mine2025}, PdTe$_{2}$ \cite{Fei2017,Bahramy2018,Clark2018,Clark2019,Salis2021,McFarane2026}, and PdPb$_{2}$ \cite{Pattalwar1988,Arushi2021,Sharma2022} crystalize in centrosymmetric structures with distinct electronic characteristics. These variations therefore offer a unique opportunity to examine how magnetotransport evolves across different electronic structuret while maintaining a common Pd-based framework. In particular, they enable a systematic comparison of the effects of crystal symmetry, spin-orbit coupling, disorder, and carrier mobility on $MR$. Furthermore, they allow us to test whether a unified transport scaling law can hold true across both centrosymmetric and noncentrosymmetric systems. 

Despite these differences, systematic magnetotransport measurements on high-purity single crystals remain limited. For these materials, previous studies have mostly focused on low-$RRR$samples, making it difficult to clear up whether the magnetotransport in these multiband, high-carrier-density systems is governed primarily by carrier compensation, or by more general parameters such as mobility and scattering in the clean limit.

Here we report a comparative magnetotransport study of $\alpha$-PdBi, $\beta$-PdBi$_{2}$, PdTe$_{2}$, and PdPb$_{2}$ using single crystals spanning a wide range of residual resistivity ratios ($RRR$), including an $\alpha$-PdBi crystal with a record-high $RRR$ of 660. By examining the field, temperature, and disorder dependence of $MR$, we identify a universal scaling behavior governed by carrier mobility, providing a unified description of magnetotransport in these complex multiband systems. The observed $MR$ exhibits a common, sub-quadratic power-law dependence on both magnetic field and $RRR$with an exponent of around 1.2 to 1.3, which clearly deviates from the quadratic behavior ($n$ = 2) of ideal compensated systems. At the same time, the centrosymmetric and noncentrosymmetric systems are clearly separated by their scaling prefactors. Furthermore, we show that high-quality crystals obey Kohler's rule despite their complex multiband Fermi surfaces. These results demonstrate that an effective single-scattering-time description works well in carrier-rich multiband metals with high purity, show the strong effect of the electronic structure over simple disorder metrics, and establish a new framework for understanding large magnetoresistance beyond the conventional compensated-semimetal scenario.

\begin{figure*}[t]
\centering
\includegraphics[width=14.8cm,clip]{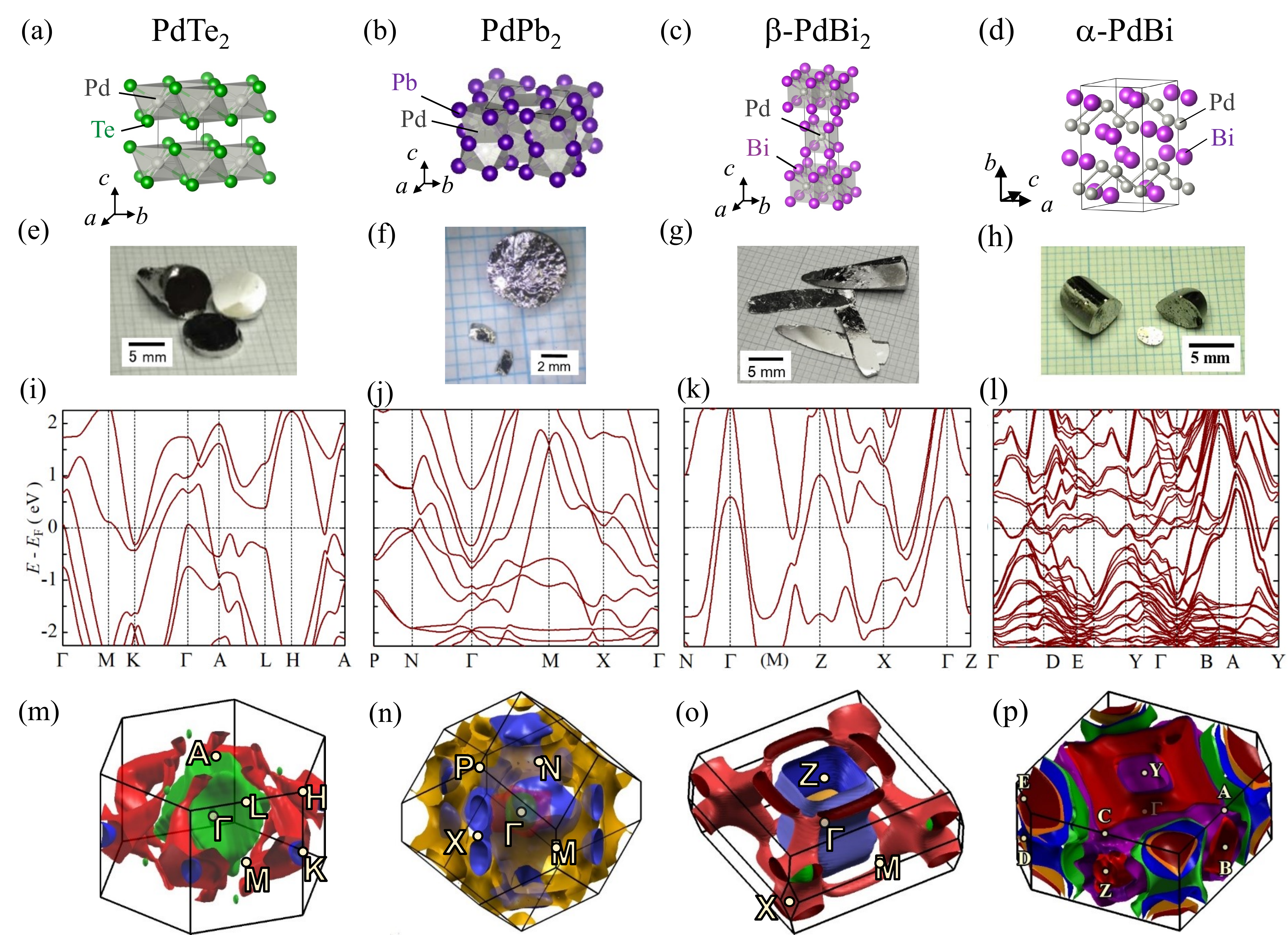}
\caption{\label{fig:wide} Crystal structure and electronic structures of Pd$X_{2}$ ($X$ = Bi, Te, Pb) and $\alpha$-PdBi. (a)--(d) Crystal structures, (e)--(h) photographs of the grown single crystals, and (i)--(l) calculated electronic band structures including spin-orbit coupling for PdTe$_{2}$, PdPb$_{2}$, $\beta$-PdBi$_{2}$, and $\alpha$-PdBi, respectively. (m)--(p) Corresponding Fermi surfaces plotted within the  first Brillouin zone, with high-symmetry points labeled using standard notation.}
\label{fig1}
\end{figure*}

\section{SINGLE CRYSTAL GROWTH}
Single crystals of Pd-based compounds, Pd$X_{2}$($X$ = Te, Pb, Bi) and $\alpha$-PdBi, were prepared by various melt-and-solution growth techniques optimized for each compound. In all cases, starting materials were sealed in evacuated quartz tubes, fully melted to ensure homogenization, and subsequently subjected to controlled cooling processes.
\subsection{Pd$X_{2}$($X$ = Te, Pb, Bi)}
For PdTe$_{2}$, a NaCl-flux method was employed to improve crystal quality. Pd, Te, and NaCl were mixed at a molar ratio of $1:2:10$, sealed in an evacuated quartz tube, and pre-reacted at high temperature. The mixture was then heated to 950~$^\circ\mathrm{C}$, held for \SI{20}{h}, and slowly cooled at a rate of 3~$^\circ\mathrm{C}\,\mathrm{h}^{-1}$ to 650~$^\circ\mathrm{C}$. Under optimized conditions\cite{Bahramy2018,Clark2018,Clark2019}, high-quality plate-like single crystals with good cleavage and typical lateral dimensions of $\sim 5\times5~\mathrm{mm}^2$ were obtained (Fig.~1(e)). 

PdPb$_{2}$ crystals were grown by a conventional melt growth technique (Pd:Pb mixture = 1:2) following the same temperature profile as PdTe$_{2}$, yielding similarly well-formed crystals (Fig. 1(f)).

Single crystals of $\beta$-PdBi$_{2}$ were also grown by a melt growth method; however, in contrast to PdPb$_{2}$, the process incorporated a rapid quenching step\cite{Sakano2015,Iwaya2017,Mine2025}. The melting mixture of Pd:Bi$ = (1:2)$ sealed in an evacuated quartz tube was held at $900~^\circ\mathrm{C}$ for \SI{20}{h}, then cooled at 3~$^\circ\mathrm{C}\,\mathrm{h}^{-1}$ to $600~^\circ\mathrm{C}$--$450~^\circ\mathrm{C}$, followed by rapid quenching into cold water. Insufficient quenching rates occasionally resulted in the formation of trace $\alpha$-PdBi$_{2}$ phase\cite{Mitra2017}. Under optimized conditions, large plate-like crystals with good cleavage and sizes up to ~$\sim1\times0.5~\mathrm{cm}^2$ were obtained (Fig.~1(g)).
\subsection{\texorpdfstring{$\alpha$}{alpha}-PdBi}
Single crystals of $\alpha$-PdBi were grown using a modified Bridgman technique\cite{Okawa2013}. Stoichiometric amounts of Pd and Bi $(1:1)$ were sealed in an evacuated quartz tube and pre-reacted until complete melting. The obtained ingot was placed in a two-zone furnace, heated to $900~^\circ\mathrm{C}$, and held for \SI{20}{h}. A temperature gradient of approximately $150~^\circ\mathrm{C}$ was then applied across the ampoule, and crystal growth was achieved by cooling both zones to room temperature at a rate of 4.5~$^\circ\mathrm{C}\,\mathrm{h}^{-1}$ while maintaining the gradient. The resulting single-grain crystals easily obtained by cleavage, exhibited shiny flat surfaces with typical dimensions of ~$\sim3\times3~\mathrm{mm}^2$ (Fig.~1(h)).

All obtained crystals have excellent cleavage properties, consistent with their layered crystal structures shown in Figs.~1(a)-1(d), and they define the current and magnetic-field directions used in subsequent transport measurements.
\begin{figure*}
\centering
\includegraphics[width=14.3cm,clip]{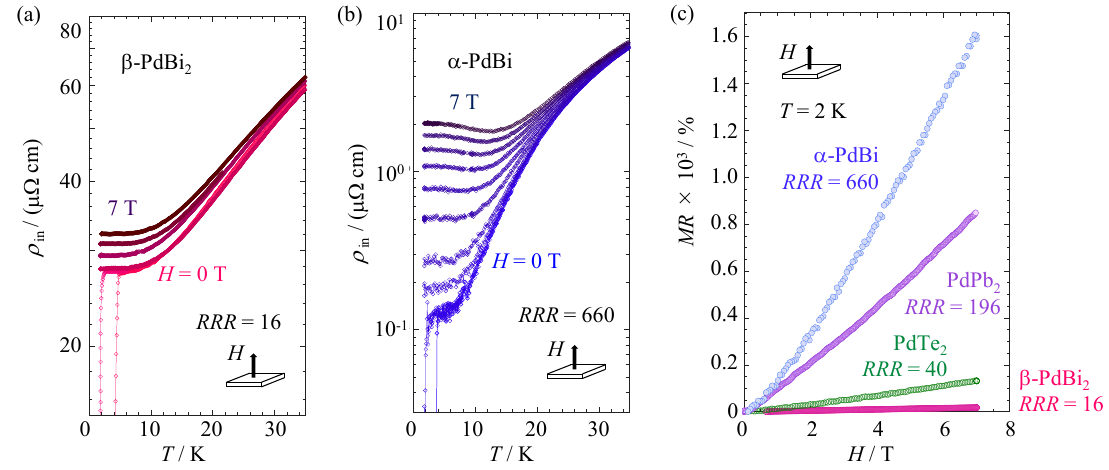}
\caption{\label{fig:wide2}Magnetotransport properties of Pd$X_{2}$ ($X$ = Bi, Te, Pb) and $\alpha$-PdBi. (a)Temperature dependence of the in-plane resistivity $\rho_{\rm in}$ in (a) $\beta$-PdBi$_{2}$ ($RRR = 16$) and (b) $\alpha$-PdBi ($RRR = 660$) single crystals under applied magnetic fields between 0 and \SI{7}{T}. The electric current was applied along the cleavage plane with the magnetic field perpendicular to it. (c) Field dependence of the magnetoresistance ratio ($MR$) at \SI{2}{K} for $\alpha$-PdBi ($RRR = 660$), PdPb$_{2}$ ($RRR = 196$), PdTe$_{2}$ ($RRR = 40$), and $\beta$-PdBi$_{2}$ ($RRR = 16$).}
\label{fig2}
\end{figure*}

\section{ELECTRICAL STRUCTURES}
First-principles calculations were performed using the WIEN2k package based on the full-potential linearized augmented plane-wave (FP-LAPW) method within the GGA-PBE approximation\cite{Blaha2018}, with experimental lattice parameters. Spin-orbit coupling (SOC) was included in a second-variational scheme. The Brillouin zone was sampled using a dense Monkhorst-Pack mesh with a momentum spacing finer than $0.03~\text{\AA}^{-1}$ for self-consistent-field calculations.

We briefly summarize the electronic structures relevant to magnetotransport, as shown in Fig.~1. The calculated band structures including spin-orbit coupling (SOC) [Figs.~1(i)--1(l)] and the corresponding Fermi surfaces [Figs.~1(m)--1(p)] demonstrate that PdTe$_{2}$, PdPb$_{2}$, $\beta$-PdBi$_{2}$, and $\alpha$-PdBi all exhibit multiple bands near the Fermi level and complex three-dimensional Fermi surface topologies.

For the centrosymmetric compounds PdTe$_{2}$, PdPb$_{2}$, and $\beta$-PdBi$_{2}$ [Figs.~1(i)--1(k) and 1(m)--1(o)], SOC induces moderate band splitting near the Fermi level, while the overall electronic structures remain qualitatively similar. Their Fermi surfaces consist of multiple electron and hole pockets with relatively large volumes, reflecting their metallic and carrier-rich nature. Unlike prototypical compensated semimetals exhibiting extremely large magnetoresistance, the calculated Fermi surfaces do not suggest a simple electron-hole compensation scenario, as the total volumes of the electron and hole pockets are not perfectly balanced. Instead, transport is expected to arise from multiple conduction channels with different carrier densities and mobilities. 

In contrast, $\alpha$-PdBi [Figs.~1(l) and 1(p)] shows a more pronounced reconstruction of the electronic structure. Owing to the absence of inversion symmetry, antisymmetric SOC lifts spin degeneracy and leads to significant band splitting. This results in strongly anisotropic Fermi surfaces, with an increased number of Fermi surface sheets and more intricate topology compared to the centrosymmetric systems. The spin-split electronic structure is expected to increase the complexity of carrier transport by introducing additional conduction channels and modifying the distribution of carrier mobilities.

Across all studied compounds, the calculated band structures and Fermi surfaces reveal a consistent feature: multiple bands crossing the Fermi level and complex, three-dimensional Fermi surface topologies. These characteristics indicate that all systems are intrinsically multiband metals with relatively high carrier densities. Therefore, conventional descriptions based on simple compensated two-band models are not expected to be directly applicable.

Notably, the degree of band dispersion and Fermi surface complexity varies among materials. In particular, the presence of steeply dispersive bands and small Fermi surface pockets, especially in the lower-symmetry compound $\alpha$-PdBi, is expected to favor higher carrier mobility through reduced effective masses. Although a quantitative determination of transport effective masses is beyond the scope of the present study, the calculated electronic structures suggest a natural origin for the material-dependent baseline of carrier mobility and $RRR$.

These results provide a microscopic basis for the experimentally observed trends in magnetotransport. They suggest that variations in carrier mobility, and hence in $RRR$, are closely linked to differences in band dispersion and Fermi surface topology. At the same time, the common presence of complex multiband Fermi surfaces across all compounds makes the emergence of simple scaling behavior particularly noteworthy. As demonstrated in the following sections, magnetotransport can nevertheless be described by unified scaling relations governed primarily by carrier mobility and effective scattering time, despite substantial differences in electronic structure.

\section{MagnetoTransports}
We investigate the magnetotransport properties of Pd based compounds, PdTe$_{2}$, PdPb$_{2}$, $\beta$-PdBi$_{2}$, and $\alpha$-PdBi, with the electric current along the natural cleavage plane and the magnetic field perpendicular to it. All compounds exhibited a metallic temperature dependence of the resistivity at zero magnetic field. Upon the application of magnetic fields up to \SI{7}{T}, a positive magnetoresistance developed at low temperatures in all compounds, while the magnitude of the $MR$ strongly depended on the material system. 

\subsection{Dependence on Material Systems}
We first examine the centrosymmetric compound $\beta$-PdBi$_{2}$, which possesses the highest crystal symmetry and a comparatively simple electronic structure among the studied compounds. The residual resistivity ratio ($RRR$), defined as $\rho(\SI{300}{K})/\rho(\SI{5}{K})$, is $\sim16$, representing a moderate value for a metallic conductor. As shown in Fig.~2(a), the application of magnetic fields induces a clear enhancement of the resistivity at low temperature, and the magnetoresistance ratio, defined as $MR = (\rho(H)-\rho(0))/\rho(0)$, reaches $\sim$\SI{18}{\%} at \SI{2}{K} and \SI{7}{T}. It is noteworthy that an $MR$ of this magnitude is already remarkably large for a conventional nonmagnetic metal with high carrier density.

In contrast, $\alpha$-PdBi, although composed of the same constituent elements, possesses a lower-symmetry crystal structure and a substantially more complex electronic structure with multiple SOC-split Fermi surface sheets due to its broken space-inversion symmetry. Surprisingly, the temperature dependence of the resistivity is significantly more pronounced, while the residual resistivity itself is reduced by more than an order of magnitude compared to $\beta$-PdBi$_{2}$. As a result, the $RRR$ increases dramatically. In the present study, we obtained $\alpha$-PdBi single crystals with a record-high $RRR$ of 660, significantly exceeding previously reported values \cite{Joshi2011,Peets2016}.

This striking enhancement of the $RRR$ in $\alpha$-PdBi cannot be explained solely by differences in crystal quality, since both $\beta$-PdBi$_{2}$ crystals and $\alpha$-PdBi crystals were synthesized using source materials of comparable purity and carefully optimized growth procedures. In metallic systems, the $RRR$ generally serves as a sensitive indicator of the clean limit of charge transport. The low-temperature residual resistivity $\rho_{0}$ is determined by the carrier mobility, reflecting both extrinsic perturbation processes such as the scattering time and intrinsic electronic structure effects such as the effective mass. Therefore, a large $RRR$ reflects not only reduced disorder scattering but also an electronic structure that is inherently favorable for high-mobility transport. 

As shown in Fig.~2(b), the temperature dependence of the in-plane resistivity $\rho_{\rm in}$ under magnetic fields reveals a pronounced enhancement of the low-temperature resistivity. At low temperature, a clear field-induced upturn appears, reflecting the emergence of an extremely large positive $MR$. For the highest-quality crystal with $RRR = 660$, the $MR$ reaches $\sim1500\%$ at \SI{2}{K} and \SI{7}{T}. These results immediately suggest a close correlation between the $MR$ and $RRR$. 

To directly compare the magnetotransport among different Pd-based metals, Fig.~2(c) summarizes the magnetic-field dependence of the $MR$ at \SI{2}{K} for $\alpha$-PdBi, PdPb$_{2}$, PdTe$_{2}$, and $\beta$-PdBi$_{2}$. The curves for PdPb$_{2}$ ($RRR = 196$) and PdTe$_{2}$ ($RRR$ = 40) are located between those of $\alpha$-PdBi ($RRR$ = 660) and $\beta$-PdBi$_{2}$ ($RRR$ = 16), demonstrating that the magnitude of the $MR$ systematically correlates with the $RRR$ across different material systems. 

\subsection{Dependence on Sample Quality in $\alpha$-PdBi}
To clarify the role of sample quality within the same material system, we systematically investigated the influence of intentionally enhanced impurity scattering in $\alpha$-PdBi. 

Figure 3(a) shows the field dependence of the $MR$ at \SI{5}{K} for $\alpha$-PdBi crystals with various $RRR$ values. The magnitude of the $MR$ decreases dramatically with decreasing $RRR$, from $\sim1500\%$ for $RRR$ = 660 to $\sim110\%$ for $RRR$ = 98. These results clearly demonstrate that the $MR$ in $\alpha$-PdBi is strongly governed by sample quality. 

The field dependence of the $MR$ can be well described by a power-law relation, 
\begin{eqnarray}
MR
= A \times H^{a}.
\label{eq1}
\end{eqnarray}
For the highest-quality crystal ($RRR = 660$), the fitting yields $a = 1.26$, indicating a sub-quadratic but non-saturating field dependence. Figure~3(b) summarizes the fitting parameters $A$ and $a$ for $\alpha$-PdBi crystals with different $RRR$ values. The prefactor $A$ strongly depends on the $RRR$, indicating that the $MR$ magnitude is highly sensitive to sample quality. In contrast, the exponent $a$ remains nearly constant at approximately 1.3 over the entire $RRR$ range. This indicates that the field dependence itself is governed by a common intrinsic mechanism that is largely unaffected by disorder scattering. Such behavior is reminiscent of compensated semimetals exhibiting extremely large $MR$ \cite{Ali2015,Chen2001,Li2006,Okawa2018}, although the present exponent is substantially smaller than the nearly quadratic behavior ($a \sim 2$) typically observed in ideal compensated systems. This difference suggests that the magnetotransport in $\alpha$-PdBi cannot be understood within a simple compensated two-band picture, despite its extremely large $MR$. 

\begin{figure}[t]
\centering
\includegraphics[width=8.32cm,clip]{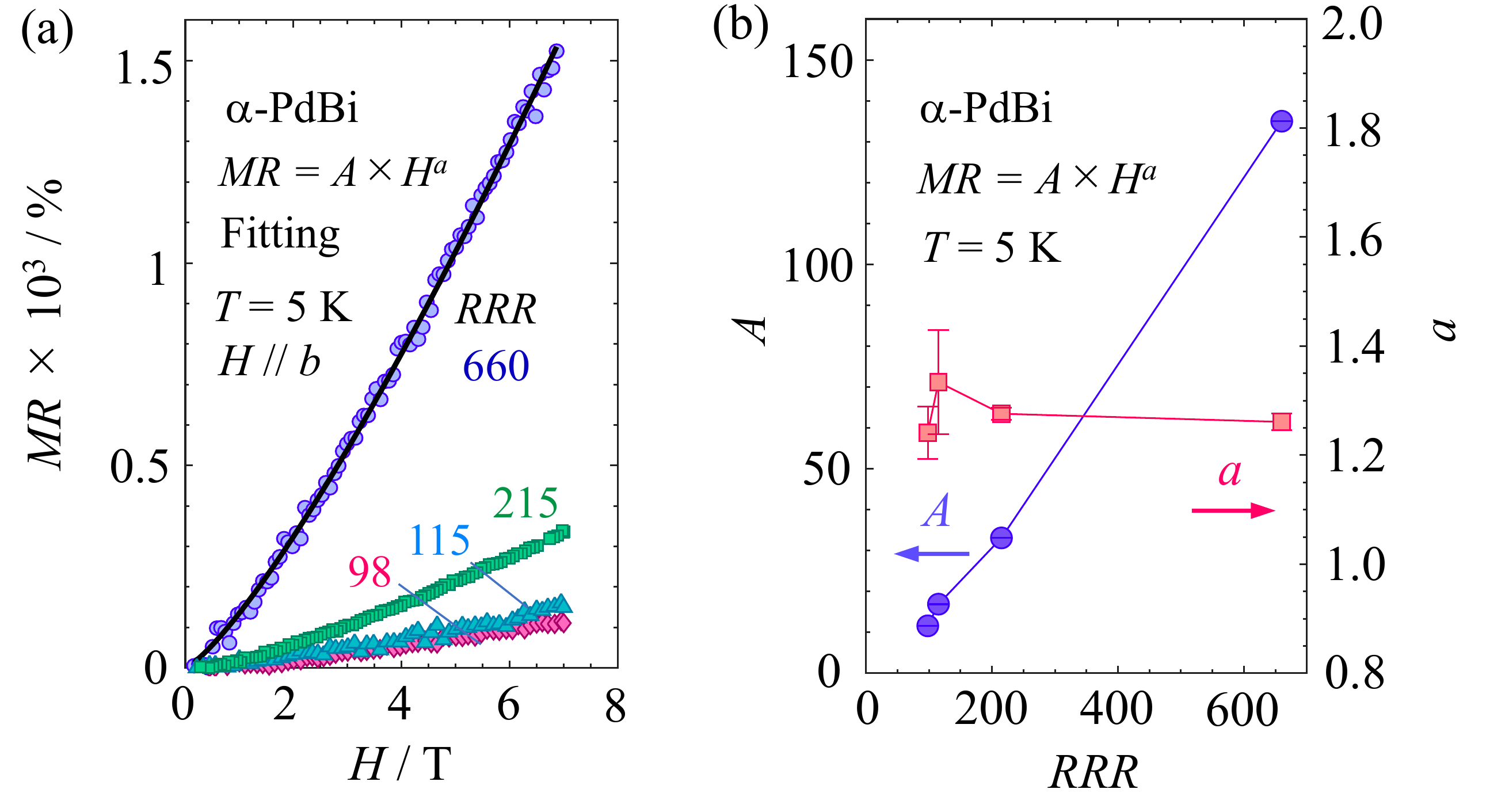}
\caption{\label{fig:wide3} $RRR$ dependence of magnetotransport properties of $\alpha$-PdBi. (a) Field dependence of the magnetoresistance ratio ($MR$) in $\alpha$-PdBi single crystals with various residual resistivity ratios ($RRR$) measured at \SI{5}{K} with the current within the cleavage plane and the applied field perpendicular to it. The black solid line represents a power-law fit, $MR = A \times H^{a}$, for the sample with $RRR$ = 660, yielding $a = 1.26$. (b) $RRR$ dependence of the power-law fitting parameters $A$ (right axis, blue symbols) and $a$ (left axis, red symbols) obtained from the relation $MR = A \times H^{a}$ for $\alpha$-PdBi.}
\label{fig3}
\end{figure}

\begin{figure*}[t]
\centering
\includegraphics[width=17.8cm,clip]{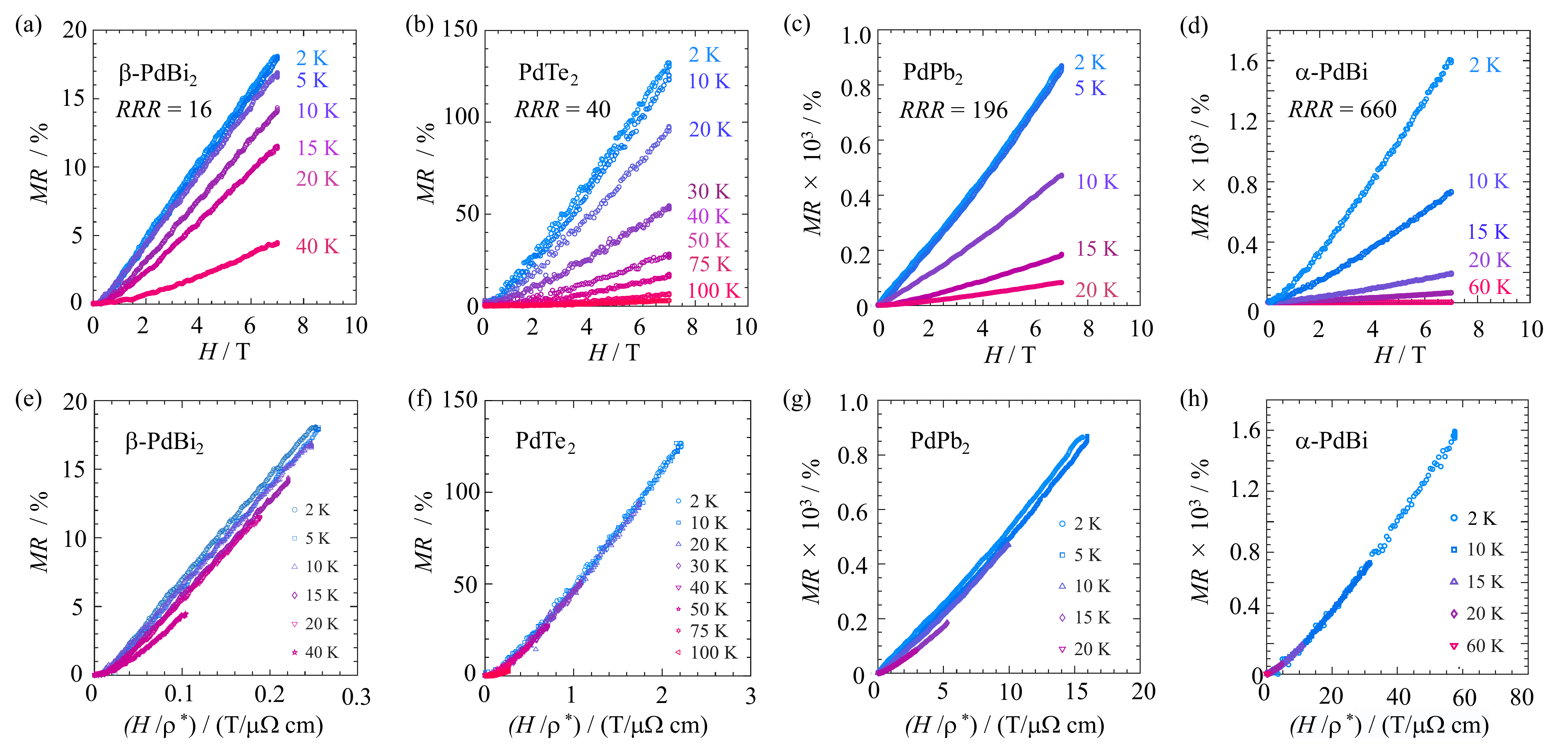}
\caption{\label{fig:wide4}
Magnetoresistance and scaling analysis of Pd$X_{2}$ ($X$ = Bi, Te, Pb)  and $\alpha$-PdBi. Magnetic field dependence of the magnetoresistance ratio ($MR$) measured at various temperatures between 2 and \SI{100}{K} for (a) $\beta$ PdBi$_{2}$ ($RRR = 16$), (b) PdTe$_{2}$ ($RRR = 40$), (c) PdPb$_{2}$ ($RRR = 196$), and (d) $\alpha$-PdBi ($RRR = 660$). The materials are arranged from left to right in order of increasing $RRR$. (e)-(h) Corresponding Kohler plots, where $MR$ is plotted as a function of the reduced field $H/\rho^{*}$, with $\rho^{*}$ being the zero-field resistivity at each temperature. 
}
\label{fig4}
\end{figure*}

\subsection{Dependence on Temperature: Kohler Scaling}
Figures~4(a)-(d) present the magnetic-field dependence of the $MR$ over a wide temperature range ($2 \sim \SI{100}{K}$) for the Pd-based metals. The materials are arranged from left to right in order of increasing $RRR$. In all compounds, the $MR$ remains positive and non-saturating up to \SI{7}{T}, although its magnitude decreases systematically with increasing temperature.

At first glance, the influence of increasing temperature resembles the reduction of the $MR$ induced by decreasing the $RRR$, as observed in Fig.~3. Since both effects suppress the $MR$ in a qualitatively similar manner, this immediately suggests that a common parameter, namely, the scattering time, plays a dominant role in governing the magnetotransport. 

To examine this possibility, we analyze the $MR$ using Kohler’s rule\cite{Kohler1938}, which is expected to hold when transport is governed by a single scattering time $\tau$ (or equivalently, a uniform carrier mobility $\mu = e\tau/m^{*}$, where $e$ and $m^{*}$ are the unit charge and the effective mass, respectively). In the weak-field limit, the $MR$ is empirically expressed as a function of the product $\omega_{\rm c}\tau$, where $\omega_{\rm c} = eH/m^{*}$ is the cyclotron frequency. If all carriers share a common scattering time, then $\tau \propto 1/\rho^{*}$ ($T,H = 0$), and the $MR$ becomes a universal function of $H/\rho^{*}$, where $\rho^{*}$ is the resistivity at a given temperature without magnetic field. Consequently, all $MR$ curves measured at different temperatures are expected to collapse onto a single scaling curve. 

The corresponding Kohler plots are shown in Figs.~4(e)-4(h). The lower-$RRR$ compound $\beta$-PdBi$_{2}$ exhibits a breakdown of Kohler scaling, suggesting the presence of multicarrier transport and/or multiple relevant scattering times \cite{Feng2025,Xu2021Kohler,Pate2024,Sonika2025}. Considering the electronic structure revealed by our first-principles calculations, where several relatively large Fermi surface sheets contribute to transport, such behavior is not unexpected for a multiband metal. 

In contrast, compounds with larger $RRR$ values exhibit remarkably good scaling behavior over a wide temperature range. Most notably, despite the absence of inversion symmetry, $\alpha$-PdBi exhibits excellent Kohler scaling. The collapse of the $MR$ data onto a single curve indicates that the magnetotransport is governed by an effective single scattering-time scale, even in the presence of SOC-split multiband Fermi surfaces. Such behavior is atypical for conventional multiband metals, where strong deviations from Kohler's rule are commonly observed\cite{Chen2025Ni}. 

An additional intriguing feature emerges from the fitting of the scaled curves. Although the prefactor $A$ differs substantially among PdTe$_{2}$, PdPb$_{2}$, and $\alpha$-PdBi, the exponent a extracted from a power-law fit [analogous to Eq.~(1)] converges to a nearly universal value of approximately 1.2--1.3. This result suggests that, once the systems enter the clean limit with sufficiently large $RRR$, the detailed multiband electronic structures become less important, and the $MR$ is governed predominantly by a common effective scattering mechanism. 

\subsection{Unified Scaling by $RRR$}
The universal trends observed in the field and temperature dependences of the $MR$ strongly suggest that the magnetotransport is governed by a common scattering-time scale across different compounds. Motivated by this observation, we further investigate whether a unified scaling relation emerges when the $MR$ is directly compared with the $RRR$, which experimentally reflects the transport lifetime.
Figure~5 summarizes the relationship between the $MR$ (measured at \SI{2}{K} and \SI{7}{T}) and the $RRR$ on a log-log scale for $\alpha$-PdBi, $\beta$-PdBi$_{2}$, PdTe$_{2}$, and PdPb$_{2}$. The data are well described by the power-law relation,
\begin{eqnarray}
MR= B \times RRR^{b}.
\label{eq2}
\end{eqnarray}
Interestingly, the plotted data separate into two distinct categories corresponding to centrosymmetric (CS) and noncentrosymmetric (NCS) compounds. Nevertheless, the fitting yields $b_{\rm CS} = 1.30~\pm~0.17$ for the CS systems and $b_{\rm NCS} = 1.18~\pm~0.11$ for the NCS system. Within experimental uncertainty, these exponents are essentially identical, indicating that the $MR-RRR$ scaling exponent is a robust and nearly universal feature across these Pd-based compounds. 

This striking universality suggests that the enhancement of the $MR$ with increasing sample quality is governed by similar scattering-controlled transport processes, largely independent of the detailed electronic structures of individual compounds. At the same time, the separation into two categories originates primarily from differences in the prefactor $B$, with the centrosymmetric compounds systematically exhibiting larger $MR$ amplitudes than the noncentrosymmetric counterpart at equivalent $RRR$ values. 

\section{discussion}
Our findings demonstrate that magnetotransport in layered Pd-based compounds, despite their multiband electronic structures, follows a remarkably simple and unified scaling behavior, once the systems enter the clean limit with sufficiently large $RRR$. This behavior can be rationalized within the semiclassical Boltzmann transport framework. 

In the case of an ideal compensated semimetal, the two-band model predicts that the magnetoresistance ($MR$) follows a quadratic field dependence, $MR \propto(\mu_{\rm ave}\times H)^{2}$, where $\mu_{\rm ave}$ represents the average carrier mobility. Given that the residual resistivity ratio ($RRR$) is approximately proportional to  $\mu_{\rm ave}$, the scaling relation becomes $MR \propto(RRR\times H)^{2}$ as exemplified by prototypical compensated semimetals including WTe$_{2}$\cite{Ali2014,Okazaki2026} and InBi\cite{Okawa2018}. 
\begin{figure}[t]
\centering
\includegraphics[width=7cm,clip]{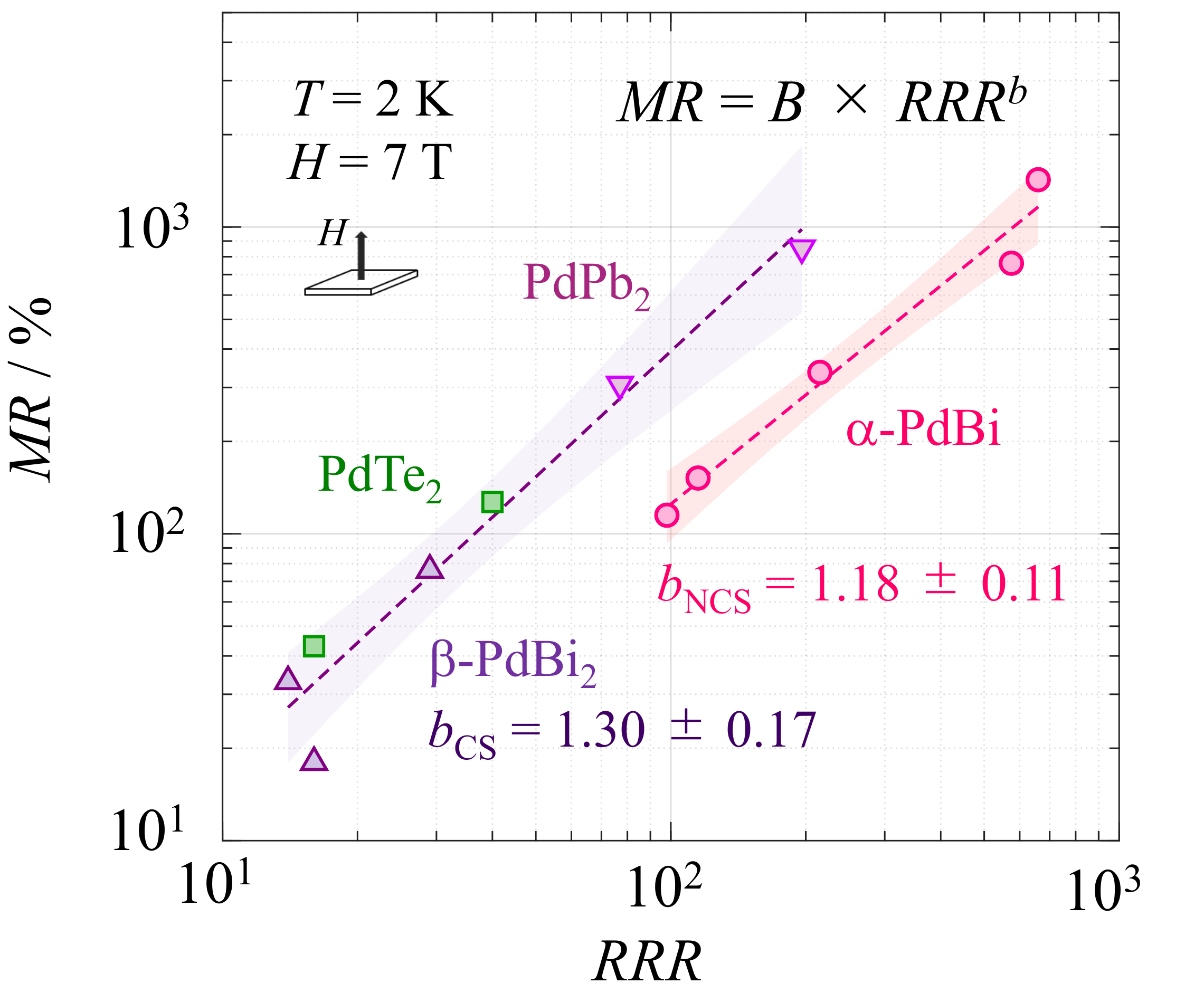}
\caption{\label{fig:wide5} Unified scaling Relationship between magnetoresistance ratio and residual resistivity ratio in Pd-based metals. Magnetoresistance ratio ($MR$) at \SI{2}{K} and \SI{7}{T} plotted as a function of the residual resistivity ratio ($RRR$) on a log-log scale for PdTe$_{2}$, PdPb$_{2}$, $\beta$-PdBi$_{2}$, and $\alpha$-PdBi single crystals. Symbol colors correspond to each material. Dotted lines represent power-law fits, $MR = B \times RRR^{b}$ for the centrosymmetric and noncentrosymmetric groups. 
} 
\label{fig5}
\end{figure}
In contrast, the present Pd-based compounds are neither perfectly compensated nor describable within a simple two-band picture. In a system with incomplete carrier compensation and multiple conduction channels, the $MR$ can be approximated as
\begin{eqnarray}
MR \sim \frac{(\mu_{\rm ave}\times H)^{2}}{1+(\Delta N/N)^{2}(\mu_{\rm ave}\times H)^{2}},
\label{eq3}
\end{eqnarray}
where $\Delta N/N$ represents the degree of imbalance between electron and hole carriers. This expression effectively captures the crossover from quadratic to nearly linear field dependence at high fields, resulting in an effective power-law behavior, 
\begin{eqnarray}
MR \propto (\mu_{\rm ave} \times H)^{n},  (1<n<2).
\label{eq4}
\end{eqnarray}
The observed intermediate exponent of $a$ in Eq.~(1) thus directly reflects the combined effects of imperfect carrier compensation and the distribution of mobilities inherent to multiband systems.

The scaling with $RRR$ can be naturally incorporated into this framework. If the dominant variation between samples arises from the scattering time, the mobility scales approximately with $RRR$, leading to 
\begin{eqnarray}
MR \propto (RRR \times H)^{n}, (1<n<2).
\label{eq5}
\end{eqnarray}

The close correspondence between the field exponent, $a$ in Eq.~(1), and the $RRR$ exponent, $b$ in Eq.~(2), observed experimentally confirms that both scaling behaviors originate from the same underlying transport mechanism. In this context, the empirical $MR$-$RRR$ scaling captures how disorder-controlled mobility governs the magnetoresponse across different samples and materials.

The validity of Kohler’s rule in high-quality crystals provides further evidence for this interpretation. The collapse of $MR$ data onto a universal curve when plotted against $H/\rho(T,H=0)$ implies that the transport can be described by an effective single scattering time scale. This observation suggests that, despite the presence of multiple bands, either transport is dominated by a subset of high-mobility carriers, or different bands share a common scattering time in the clean limit. The negligible temperature-dependent deviations further imply that variations in carrier density and effective mass are not the primary drivers of the observed scaling behavior.

An intriguing aspect of the present results is that the scaling relation Eq.~(2) is not only observed within a given compound but also persists across multiple centrosymmetric materials ($\beta$-PdBi$_{2}$, PdTe$_{2}$, and PdPb$_{2}$). This indicates that the underlying transport mechanism is robust against changes in the detailed band structure, pointing to a universal description governed by mobility and scattering rather than material-specific fine structure.

At the same time, the overall magnitude of the $MR$ exhibits clear material dependence. Within a simplified transport picture, the mobility is given by $\mu_{\rm ave} \sim \tau/m^{*}$, suggesting that both scattering time and effective mass contribute to the observed differences. Given that the crystals were grown under carefully optimized and comparable conditions, it is reasonable to assume that variations in impurity scattering are not the dominant source of the differences in $RRR$ between materials. Instead, the effective mass, reflecting the each electronic structure, is expected to play a key role.

First-principles calculations indeed reveal notable variations in band dispersion and Fermi surface topology among the studied compounds. In particular, the presence of steeply dispersive bands and small Fermi surface pockets is expected to enhance carrier mobility by reducing the effective mass, providing a physical basis for the material-dependent $RRR$ baseline.

Finally, the comparison between centrosymmetric and noncentrosymmetric compounds highlights the role of inversion symmetry. Although $\alpha$-PdBi exhibits the largest absolute $MR$ due to its exceptionally high $RRR$, its magnetotransport efficiency is suppressed compared to centrosymmetric systems at equivalent $RRR$ values ($B_{\rm NCS} < B_{\rm CS}$ in Eq.~(2)). In noncentrosymmetric systems, spin-orbit coupling lifts spin degeneracy through Rashba-type band splitting, effectively increasing the number of conduction channels and introducing additional scattering pathways. These effects reduce the effective carrier mobility and lead to a suppression of the $MR$ compared to centrosymmetric counterparts.

Altogether, these results demonstrate that magnetotransport in Pd-based multiband metals can be understood within a unified framework combining disorder-controlled mobility, imperfect carrier compensation, and band-dependent transport. This framework naturally explains the emergence of intermediate power-law scaling and its robustness across different materials, providing a new perspective on large magnetoresistance beyond the conventional compensated semimetal scenario.

\section{Conclusions}
In conclusion, we have performed a comprehensive study of magnetoresistance in single-crystalline, nonmagnetic layered Pd-based metals, unveiling a universal scaling behavior that persists across materials with distinct electronic structures. Despite their multiband character and relatively high carrier densities, the magnetoresistance ratio ($MR$) displays remarkably robust and consistent dependencies on magnetic field, temperature, and sample quality. 

A key outcome of this work is the clear correlation between $MR$ and the residual resistivity ratio ($RRR$), demonstrating that carrier mobility plays a dominant role in determining the transport response. The validity of Kohler scaling in high-$RRR$ crystals indicates that carrier transport can be described by an effective single scattering time scale, even in the presence of multiple Fermi surface sheets. The observed sub-quadratic field and $RRR$ dependencies further reflect the combined influence of incomplete carrier compensation and the coexistence of carriers with different mobilities. 

Although sample-to-sample variations in $RRR$ arise primarily from disorder, systematic differences between compounds suggest an intrinsic contribution originating from the underlying electronic structure. First-principles calculations reveal notable variations in band dispersion and Fermi surface topology among the studied materials. In particular, the presence of steeply dispersive bands and small Fermi surface pockets is expected to favor higher carrier mobility, providing a qualitative account of the intrinsic $RRR$ limits across different compounds. 

A comparison between centrosymmetric and noncentrosymmetric compounds reveals that inversion symmetry governs $MR$ in a nontrivial manner. While $\alpha$-PdBi exhibits the largest absolute $MR$ due to its exceptionally high $RRR$, its response is systematically suppressed compared to centrosymmetric systems at equivalent $RRR$ values. This behavior suggests that spin-orbit-induced band splitting in noncentrosymmetric systems introduces additional scattering channels, which effectively limit carrier mobility. 

Overall, the present results demonstrate that magnetotransport in these Pd-based metals is governed by a cooperative interplay between disorder, band-dependent mobility, and symmetry. This work provides a framework for understanding large $MR$ beyond the conventional compensated semimetal scenario and offers guiding principles for exploring high-mobility transport in complex multiband systems.

\begin{acknowledgments}
This work was supported by a research fellowship for young scientists (Grants No. 15J11876), a Grant-in-Aid (KAKENHI) for Early-Career Scientists (Grant No.23K13552), for Scientific Research (C) (Grant No.23K13552), for Scientific Research (A) (Grants Nos. JP21H04652 and JP26H02235), for Challenging Research (Pioneering) (JP21K18181), and for Scientific Research on Innovative Areas (JP26H00647) from the Japan Society for Promotion of Science. This work was also supported by the Thermal \& Electric Energy Technology Foundation (TEET), Iketani Science and Technology Foundation, and the Precise Measurement Technology Promotion Foundation (PMTP-F). 
\end{acknowledgments}

\bibliography{Pd-MR}%

@article{Gao2020,
  author = {Gao, J. J. and Si, J. G. and Luo, X. and Yan, J. and Jiang, Z. Z. and Wang, W. and Han, Y. Y. and Tong, P. and Song, W. H. and Zhu, X. B. and Li, Q. J. and Lu, W. J. and Sun, Y. P.},
  title = {Origin of the large magnetoresistance in the candidate chiral superconductor {$4H_b$-TaS$_2$}},
  journal = {Phys. Rev. B},
  volume = {102},
  issue = {7},
  pages = {075138},
  year = {2020},
  doi = {10.1103/PhysRevB.102.075138}
}

@article{Sundar2019,
  author = {Sundar, S. and Salem-Sugui, S. Jr. and Chattopadhyay, M. K. and Roy, S. B. and Sharath Chandra, L. S. and Cohen, L. F. and Ghivelder, L.},
  title = {Study of {Nb$_{0.18}$Re$_{0.82}$} non-centrosymmetric superconductor in the normal and superconducting states},
  journal = {Supercond. Sci. Technol.},
  volume = {32},
  pages = {055003},
  year = {2019},
  doi = {10.1088/1361-6668/ab06a5}
}

@article{Mu2010,
  author = {Mu, G. and Huang, H. and Wen, H.-H.},
  title = {Multiband effect in the noncentrosymmetric superconductor {Mg$_{12-\delta}$Ir$_{19}$B$_{16}$} revealed by Hall effect and magnetoresistance measurements},
  journal = {Phys. Rev. B},
  volume = {82},
  pages = {052501},
  year = {2010},
  doi = {10.1103/PhysRevB.82.052501}
}

@article{Xing2012,
  author = {Xing, J. and Li, S. and Ding, X. and Yang, H. and Wen, H.-H.},
  title = {Superconductivity appears in the vicinity of semiconducting-like behavior in {CeO$_{1-x}$F$_x$BiS$_2$}},
  journal = {Phys. Rev. B},
  volume = {86},
  pages = {214518},
  year = {2012},
  doi = {10.1103/PhysRevB.86.214518}
}

@article{Li2013,
  author = {Li, S. and Yang, H. and Fang, D.},
  title = {Strong coupling superconductivity and prominent superconducting fluctuations in the new superconductor {Bi$_4$O$_4$S$_3$}},
  journal = {Sci. China Phys. Mech. Astron.},
  volume = {56},
  pages = {2019},
  year = {2013},
  doi = {10.1007/s11433-013-5323-y}
}

@article{Sun2014,
  author = {Sun, Y. and Taen, T. and Yamada, T. and Pyon, S. and Nishizaki, T. and Shi, Z. and Tamegai, T.},
  title = {Multiband effects and possible Dirac fermions in {Fe$_{1+y}$Te$_{0.6}$Se$_{0.4}$}},
  journal = {Phys. Rev. B},
  volume = {89},
  pages = {144512},
  year = {2014},
  doi = {10.1103/PhysRevB.89.144512}
}

@article{Lin2016,
  author = {Lin, H. and Li, Y. and Deng, Q. and Xing, J. and Liu, J. and Zhu, X. and Yang, H. and Wen, H.-H.},
  title = {Multiband superconductivity and large anisotropy in {FeS} crystals},
  journal = {Phys. Rev. B},
  volume = {93},
  pages = {144512},
  year = {2016},
  doi = {10.1103/PhysRevB.93.144512}
}

@article{Wang2025,
  author = {Wang, X. and Sun, Y. and Wei, W. and Hou, Q. and Zhou, N. and Zhang, Y. and Shi, Z.},
  title = {Strong correlation between {$H$}-linear magnetoresistance and strange metal in the {FeSe} superconductor},
  journal = {Phys. Rev. B},
  volume = {111},
  pages = {L100501},
  year = {2025},
  doi = {10.1103/PhysRevB.111.L100501},
  url = {https://link.aps.org/doi/10.1103/PhysRevB.111.L100501}
}

@article{Watson2015,
  author = {Watson, M. D. and Yamashita, T. and Kasahara, S. and Knafo, W. and Nardone, M. and B{\'e}ard, J. and Hardy, F. and McCollam, A. and Narayanan, A. and Blake, S. F. and Wolf, T. and Haghighirad, A. A. and Meingast, C. and Schofield, A. J. and v. L{\"o}hneysen, H. and Matsuda, Y. and Coldea, A. I. and Shibauchi, T.},
  title = {Dichotomy between the hole and electron behavior in multiband superconductor {FeSe} probed by ultrahigh magnetic fields},
  journal = {Phys. Rev. Lett.},
  volume = {115},
  pages = {027006},
  year = {2015},
  doi = {10.1103/PhysRevLett.115.027006}
}

@article{Oudau2020,
  author = {Oudah, Mohamed and Bannies, Jonas and Bonn, D. A. and Aronson, M. C.},
  title = {Superconductivity and quantum oscillations in single crystals of the compensated semimetal {CaSb$_2$}},
  journal = {Phys. Rev. B},
  volume = {105},
  pages = {184504},
  year = {2022},
  doi = {10.1103/PhysRevB.105.184504}
}

@article{Funada2019,
  author = {Funada, K. and Yamakage, A. and Yamashita, N. and Kageyama, H.},
  title = {Spin-Orbit-Coupling-Induced Type-I/Type-II Dirac Nodal-Line Metal in Nonsymmorphic {CaSb$_2$}},
  journal = {J. Phys. Soc. Jpn.},
  volume = {88},
  pages = {044711},
  year = {2019},
  doi = {10.7566/JPSJ.88.044711}
}

@article{Rhodes2017,
  author = {Rhodes, D. and Sch{\"o}nemann, R. and Aryal, N. and Zhou, Q. and Zhang, Q. R. and Kampert, E. and Chiu, Y.-C. and Lai, Y. and Shimura, Y. and McCandless, G. T. and Chan, J. Y. and Paley, D. W. and Lee, J. and Finke, A. D. and Ruff, J. P. C. and Das, S. and Manousakis, E. and Balicas, L.},
  title = {Bulk Fermi surface of the Weyl type-II semimetallic candidate {$\gamma$-MoTe$_2$}},
  journal = {Phys. Rev. B},
  volume = {96},
  pages = {165134},
  year = {2017},
  doi = {10.1103/PhysRevB.96.165134}
}

@article{Chen2016,
  author = {Chen, F. C. and Lv, H. Y. and Luo, X. and Lu, W. J. and Pei, Q. L. and Lin, G. T. and Han, Y. Y. and Zhu, X. B. and Song, W. H. and Sun, Y. P.},
  title = {Extremely large magnetoresistance in the type-II Weyl semimetal {MoTe$_2$}},
  journal = {Phys. Rev. B},
  volume = {94},
  pages = {235154},
  year = {2016},
  doi = {10.1103/PhysRevB.94.235154}
}

@article{Okazaki2026,
  author = {Okazaki, Shota and Sasagawa, Takao},
  title = {Extremely large magnetoresistance and quantum oscillations in ultra-high-quality single crystals of the Weyl semimetal {WTe$_2$}},
  journal = {Phys. Rev. Mater.},
  volume = {10},
  pages = {L051202},
  year = {2026},
  doi = {10.1103/dnv7-4xgd}
}

@book{Pippard1989,
  author = {Pippard, A. B.},
  title = {Magnetoresistance in Metals},
  publisher = {Cambridge University Press},
  address = {Cambridge},
  year = {1989}
}

@article{Niu2021,
  author = {Niu, Rui and Zhu, W. K.},
  title = {Materials and possible mechanisms of extremely large magnetoresistance: A review},
  journal = {J. Phys.: Condens. Matter},
  volume = {34},
  number = {11},
  pages = {113001},
  year = {2022},
  doi = {10.1088/1361-648X/ac3b24}
}

@article{Hu2008,
  author = {Hu, J. and Rosenbaum, T. F.},
  title = {Classical and quantum routes to linear magnetoresistance},
  journal = {Nat. Mater.},
  volume = {7},
  pages = {697--700},
  year = {2008},
  doi = {10.1038/nmat2253}
}

@article{Parish2003,
  author = {Parish, M. M. and Littlewood, P. B.},
  title = {Non-saturating magnetoresistance in heavily disordered semiconductors},
  journal = {Nature},
  volume = {426},
  pages = {162--165},
  year = {2003},
  doi = {10.1038/nature02073}
}

@article{Abrikosov1998,
  author = {Abrikosov, A. A.},
  title = {Quantum magnetoresistance},
  journal = {Phys. Rev. B},
  volume = {58},
  pages = {2788--2794},
  year = {1998},
  doi = {10.1103/PhysRevB.58.2788}
}

@article{Ali2014,
  author = {Ali, Mazhar N. and Xiong, Jun and Flynn, Steven and Tao, Jing and Gibson, Quinn D. and Schoop, Leslie M. and Liang, Tian and Haldolaarachchige, Neel and Hirschberger, Max and Ong, N. P. and Cava, R. J.},
  title = {Large, non-saturating magnetoresistance in {WTe$_2$}},
  journal = {Nature},
  volume = {514},
  pages = {205--208},
  year = {2014},
  doi = {10.1038/nature13763}
}

@article{Ali2015,
  author = {Ali, Mazhar N. and Schoop, Leslie and Xiong, Jun and Flynn, Steven and Gibson, Quinn and Hirschberger, Max and Ong, N. P. and Cava, R. J.},
  title = {Correlation of crystal quality and extreme magnetoresistance of {WTe$_2$}},
  journal = {EPL},
  volume = {110},
  pages = {67002},
  year = {2015},
  doi = {10.1209/0295-5075/110/67002}
}

@article{Shekhar2015,
  author = {Shekhar, Chandra and Nayak, Ajaya K. and Sun, Yan and Schmidt, Marcus and Nicklas, Michael and Leermakers, Inge and Zeitler, Uli and Skourski, Yurii and Wosnitza, Jochen and Liu, Zhongkai and Chen, Yulin and Schnelle, Walter and Borrmann, Horst and Grin, Yuri and Felser, Claudia and Yan, Binghai},
  title = {Extremely large magnetoresistance and ultrahigh mobility in the topological Weyl semimetal candidate {NbP}},
  journal = {Nat. Phys.},
  volume = {11},
  pages = {645--649},
  year = {2015},
  doi = {10.1038/nphys3372}
}

@article{Leahy2018,
  author = {Leahy, Ian A. and Lin, Yu-Ping and Siegfried, Peter E. and Treglia, Andrew C. and Song, Justin C. W. and Nandkishore, Rahul M. and Lee, Minhyea},
  title = {Nonsaturating large magnetoresistance in semimetals},
  journal = {Proc. Natl. Acad. Sci. U.S.A.},
  volume = {115},
  number = {42},
  pages = {10570--10575},
  year = {2018},
  doi = {10.1073/pnas.1808747115}
}

@article{Gao2018,
  author = {Gao, Wenshuai and Zhu, Xiangde and Zheng, Fawei and Wu, Min and Zhang, Jinglei and Xi, Chuanying and Zhang, Ping and Zhang, Yuheng and Hao, Ning and Ning, Wei and Tian, Mingliang},
  title = {A possible candidate for triply degenerate point fermions in trigonal layered {PtBi$_2$}},
  journal = {Phys. Rev. B},
  volume = {98},
  pages = {121108},
  year = {2018},
  doi = {10.1038/s41467-018-05730-3}
}

@article{Wu2020,
  author = {Wu, Beilun and Barrena, V{'i}ctor and Suderow, Hermann and Guillam{'o}n, Isabel},
  title = {Huge linear magnetoresistance due to open orbits in {$\gamma$-PtBi$_2$}},
  journal = {Phys. Rev. Research},
  volume = {2},
  pages = {022042},
  year = {2020},
  doi = {10.1103/PhysRevResearch.2.022042}
}

@article{Xing2020,
  author = {Xing, Lingyi and Chapai, Ramakanta and Nepal, Roshan and Jin, Rongying},
  title = {Topological behavior and Zeeman splitting in trigonal {PtBi$_{2-x}$} single crystals},
  journal = {npj Quantum Mater.},
  volume = {5},
  pages = {10},
  year = {2020},
  doi = {10.1038/s41535-020-0213-9}

}

@article{Joshi2011,
  author = {Joshi, Bhanu and Thamizhavel, A. and Ramakrishnan, S.},
  title = {Superconductivity in noncentrosymmetric {BiPd}},
  journal = {Phys. Rev. B},
  volume = {84},
  pages = {064518},
  year = {2011},
  doi = {10.1103/PhysRevB.84.064518}
}

@article{Okawa2013,
  author = {Okawa, K. and Kanou, M. and Katagiri, T. and Kashiwaya, H. and Kashiwaya, S. and Sasagawa, T.},
  title = {Crystal growth and physical properties of the noncentrosymmetric superconductor {PdBi}},
  journal = {Phys. Procedia},
  volume = {45},
  pages = {101--104},
  year = {2013},
  doi = {10.1016/j.phpro.2013.04.062}

}

@article{Matano2013,
  author = {Matano, Kazuaki and Maeda, Satoki and Sawaoka, Hiroki and Muro, Yuji and Takabatake, Toshiro and Joshi, Bhanu and Ramakrishnan, Srinivasan and Kawashima, Kenji and Akimitsu, Jun and Zheng, Guo-qing},
  title = {{NMR} and {NQR} studies on non-centrosymmetric superconductors {Re$_7$B$_3$}, {LaBiPt}, and {BiPd}},
  journal = {J. Phys. Soc. Jpn.},
  volume = {82},
  pages = {084711},
  year = {2013},
  doi = {10.7566/JPSJ.82.084711}
}

@article{Neupane2016,
  author = {Neupane, Madhab and Alidoust, Nasser and Hosen, M. Mofazzel and Zhu, Jian-Xin and Dimitri, Klauss and Xu, Su-Yang and Dhakal, Nagendra and Sankar, Raman and Belopolski, Ilya and Sanchez, Daniel S. and Chang, Tay-Rong and Jeng, Horng-Tay and Miyamoto, Koji and Okuda, Taichi and Lin, Hsin and Bansil, Arun and Kaczorowski, Dariusz and Chou, Fangcheng and Hasan, M. Zahid and Durakiewicz, Tomasz},
  title = {Observation of the spin-polarized surface state in a noncentrosymmetric superconductor {BiPd}},
  journal = {Nat. Commun.},
  volume = {7},
  pages = {13315},
  year = {2016},
  doi = {10.1038/ncomms13315}

}

@article{Benia2016,
  author = {Benia, H. M. and Rampi, E. and Trainer, C. and Yim, C. M. and Maldonado, A. and Peets, D. C. and St{"o}hr, A. and Starke, U. and Kern, K. and Yaresko, A. and Levy, G. and Damascelli, A. and Ast, C. R. and Schnyder, A. P. and Wahl, P.},
  title = {Observation of Dirac surface states in the noncentrosymmetric superconductor {BiPd}},
  journal = {Phys. Rev. B},
  volume = {94},
  pages = {121407},
  year = {2016},
  doi = {10.1103/PhysRevB.94.121407}
}

@article{Thirupathaiah2016,
  author = {Thirupathaiah, S. and Ghosh, Soumi and Jha, Rajveer and Rienks, E. D. L. and Dolui, Kapildeb and Ravi Kishore, V. V. and B{"u}chner, B. and Das, Tanmoy and Awana, V. P. S. and Sarma, D. D. and Fink, J.},
  title = {Unusual Dirac fermions on the surface of a noncentrosymmetric {$\alpha$-BiPd} superconductor},
  journal = {Phys. Rev. Lett.},
  volume = {117},
  pages = {177001},
  year = {2016},
  doi = {10.1103/PhysRevLett.117.177001}
}

@article{Lohani2017,
  author = {Lohani, H. and Mishra, P. and Gupta, Anurag and Awana, V. P. S. and Sekhar, B. R.},
  title = {Fermi surface and band structure of {BiPd} from {ARPES} studies},
  journal = {Physica C},
  volume = {534},
  pages = {13--18},
  year = {2017},
  doi = {10.1016/j.physc.2016.12.004}

}

@article{Pramanik2021,
  author = {Pramanik, Arindam and Pandeya, Ram Prakash and Vyalikh, Denis V. and Generalov, Alexander and Moras, Paolo and Kundu, Asish K. and Sheverdyaeva, Polina M. and Carbone, Carlo and Joshi, Bhanu and Thamizhavel, A. and Ramakrishnan, S. and Maiti, Kalobaran},
  title = {Dirac states in the noncentrosymmetric superconductor {BiPd}},
  journal = {Phys. Rev. B},
  volume = {103},
  pages = {155401},
  year = {2021},
  doi = {10.1103/PhysRevB.103.155401}
}

@article{Jiao2014,
  author = {Jiao, L. and Zhang, J. L. and Chen, Y. and Weng, Z. F. and Shao, Y. M. and Feng, J. Y. and Lu, X. and Joshi, B. and Thamizhavel, A. and Yuan, H. Q.},
  title = {Anisotropic superconductivity in noncentrosymmetric {BiPd}},
  journal = {Phys. Rev. B},
  volume = {89},
  pages = {060507},
  year = {2014},
  doi = {10.1103/PhysRevB.89.060507}
}

@article{Sun2015,
  author = {Sun, Zhixiang and Enayat, Mostafa and Maldonado, Ana and Lithgow, Calum and Yelland, Ed and Peets, Darren C. and Yaresko, Alexander and Schnyder, Andreas P. and Wahl, Peter},
  title = {Dirac surface states and nature of superconductivity in noncentrosymmetric {BiPd}},
  journal = {Nat. Commun.},
  volume = {6},
  pages = {6633},
  year = {2015},
  doi = {10.1038/ncomms7633}

}

@article{Peets2016,
  author = {Peets, Darren C. and Maldonado, Ana and Enayat, Mostafa and Sun, Zhixiang and Wahl, Peter and Schnyder, Andreas P.},
  title = {Upper critical field of the noncentrosymmetric superconductor {BiPd}},
  journal = {Phys. Rev. B},
  volume = {93},
  pages = {174504},
  year = {2016},
  doi = {10.1103/PhysRevB.93.174504}
}

@article{Yim2018,
  author = {Yim, Chi Ming and Trainer, Christopher and Maldonado, Ana and Braunecker, Bernd and Yaresko, Alexander and Peets, Darren C. and Wahl, Peter},
  title = {Kinetic stabilization of {1D} surface states near twin boundaries in noncentrosymmetric {BiPd}},
  journal = {Phys. Rev. Lett.},
  volume = {121},
  pages = {206401},
  year = {2018},
  doi = {10.1103/PhysRevLett.121.206401}
}

@article{Cameron2025,
  author = {Cameron, A. S. and Lemberger, L. and Riyat, R. and Holmes, A. T. and Yerin, Y. S. and Hillier, A. D. and Joshi, B. and Ramakrishnan, S. and Gavilano, J. and Dewhurst, C. D. and Forgan, E. M. and Blackburn, E.},
  title = {Unconventional gap structures and the intermediate mixed state: A vortex lattice study of the noncentrosymmetric superconductor {BiPd}},
  journal = {Phys. Rev. B},
  volume = {111},
  pages = {094514},
  year = {2025},
  doi = {10.1103/PhysRevB.111.094514}
}

@article{Mondal2013,
  author = {Mondal, Mintu and Joshi, Bhanu and Kumar, Sanjeev and Kamlapure, Anand and Ganguli, Somesh Chandra and Thamizhavel, Arumugam and Mandal, Sudhansu S. and Ramakrishnan, Srinivasan and Raychaudhuri, Pratap},
  title = {Andreev bound state and multiple energy gaps in the noncentrosymmetric superconductor {BiPd}},
  journal = {Phys. Rev. B},
  volume = {86},
  pages = {094520},
  year = {2012},
  doi = {10.1103/PhysRevB.86.094520}
}

@article{Klotz2020,
  author = {Klotz, J. and Butcher, T. A. and F{"o}rster, T. and Hornung, J. and Sheikin, I. and Wisniewski, P. and Jesche, A. and Wosnitza, J. and Kaczorowski, D.},
  title = {Fermi surface investigation of the noncentrosymmetric superconductor {$\alpha$-PdBi}},
  journal = {Phys. Rev. B},
  volume = {101},
  pages = {235139},
  year = {2020},
  doi = {10.1103/PhysRevB.101.235139}
}

@article{Khan2019,
  author = {Khan, Mojammel A. and Graf, D. E. and Vekhter, I. and Browne, D. A. and DiTusa, J. F. and Phelan, W. Adam and Young, D. P.},
  title = {Quantum oscillations and a nontrivial Berry phase in the noncentrosymmetric topological superconductor candidate {BiPd}},
  journal = {Phys. Rev. B},
  volume = {99},
  pages = {020507},
  year = {2019},
  doi = {10.1103/PhysRevB.99.020507}
}

@article{Pramanik2020,
  author = {Pramanik, Arindam and Pandeya, Ram Prakash and Ali, Khadiza and Joshi, Bhanu and Sarkar, Indranil and Moras, Paolo and Sheverdyaeva, Polina M. and Kundu, Asish K. and Carbone, Carlo and Thamizhavel, A. and Ramakrishnan, S. and Maiti, Kalobaran},
  title = {Depth-resolved core level spectroscopy of noncentrosymmetric solid {BiPd}},
  journal = {Phys. Rev. B},
  volume = {101},
  issue = {3},
  pages = {035426},
  year = {2020},
  doi = {10.1103/PhysRevB.101.035426}
}

@article{Yaresko2018,
  author = {Yaresko, Alexander and Schnyder, Andreas P. and Benia, Hadj M. and Yim, Chi-Ming and Levy, Giorgio and Damascelli, Andrea and Ast, Christian R. and Peets, Darren C. and Wahl, Peter},
  title = {Correct Brillouin zone and electronic structure of {BiPd}},
  journal = {Phys. Rev. B},
  volume = {97},
  pages = {075108},
  year = {2018},
  doi = {10.1103/PhysRevB.97.075108}
}

@article{Imai2012,
  author = {Imai, Yoshinori and Nabeshima, Fuyuki and Yoshinaka, Taiki and Miyatani, Kosuke and Kondo, Ryusuke and Komiya, Seiki and Tsukada, Ichiro and Maeda, Atsutaka},
  title = {Superconductivity at {5.4 K} in {$\beta$-Bi$_2$Pd}},
  journal = {J. Phys. Soc. Jpn.},
  volume = {81},
  pages = {113708},
  year = {2012},
  doi = {10.1143/JPSJ.81.113708}
}

@article{Sakano2015,
  author = {Sakano, M. and Okawa, K. and Kanou, M. and Sanjo, H. and Okuda, T. and Sasagawa, T. and Ishizaka, K.},
  title = {Topologically protected surface states in a centrosymmetric superconductor {$\beta$-PdBi$_2$}},
  journal = {Nat. Commun.},
  volume = {6},
  pages = {8595},
  year = {2015},
  doi = {10.1038/ncomms9595}

}

@article{Iwaya2017,
  author = {Iwaya, K. and Kohsaka, Y. and Okawa, K. and Machida, T. and Bahramy, M. S. and Hanaguri, T. and Sasagawa, T.},
  title = {Full-gap superconductivity in spin-polarized surface states of topological semimetal {$\beta$-PdBi$_2$}},
  journal = {Nat. Commun.},
  volume = {8},
  pages = {976},
  year = {2017},
  doi = {10.1038/s41467-017-01209-9}

}

@article{Powell2025,
  author = {Powell, Lewis and Kuang, Wenjun and Hawkins-Pottier, Gabriel and Jalil, Rashid and Birkbeck, John and Jiang, Ziyi and Kim, Minsoo and Zou, Yichao and Komrakova, Sofiia and Haigh, Sarah and Timokhin, Ivan and Balakrishnan, Geetha and Geim, Andre K. and Walet, Niels and Principi, Alessandro and Grigorieva, Irina V.},
  title = {Multiphase superconductivity in {PdBi$_2$}},
  journal = {Nat. Commun.},
  volume = {16},
  pages = {291},
  year = {2025},
  doi = {10.1038/s41467-024-54867-x}

}

@article{Mine2025,
  author = {Mine, Akifumi and Suzuki, Takeshi and Zhong, Yigui and Najafzadeh, Sahand and Okawa, Kenjiro and Sakano, Masato and Ishizaka, Kyoko and Shin, Shik and Sasagawa, Takao and Okazaki, Kozo},
  title = {Direct observation of the surface superconducting gap in the topological superconductor candidate {$\beta$-PdBi$_2$}},
  journal = {Phys. Rev. Lett.},
  volume = {135},
  pages = {236002},
  year = {2025},
  doi = {10.1103/3jgn-22rx}
}

@article{Fei2017,
  author = {Fei, Fucong and Bo, Xiangyan and Wang, Rui and Wu, Bin and Jiang, Juan and Fu, Dongzhi and Gao, Ming and Zheng, Hao and Chen, Yulin and Bu, Haijun and Song, Fengqi and Wan, Xiangang and Wang, Baigeng and Wang, Guanghou},
  title = {Nontrivial Berry phase and type-II Dirac transport in the layered material {PdTe$_2$}},
  journal = {Phys. Rev. B},
  volume = {96},
  pages = {041201},
  year = {2017},
  doi = {10.1103/PhysRevB.96.041201}
}

@article{Bahramy2018,
  author = {Bahramy, M. S. and Clark, O. J. and Yang, B.-J. and Feng, J. and Bawden, L. and Riley, J. M. and Markovi{'c}, I. and Mazzola, F. and Sunko, V. and Biswas, D. and Cooil, S. P. and Jorge, M. and Wells, J. W. and Leandersson, M. and Balasubramanian, T. and Fujii, J. and Vobornik, I. and Rault, J. E. and Kim, T. K. and Hoesch, M. and Okawa, K. and Asakawa, M. and Sasagawa, T. and Eknapakul, T. and Meevasana, W. and King, P. D. C.},
  title = {Ubiquitous formation of bulk Dirac cones and topological surface states from a single orbital manifold in transition-metal dichalcogenides},
  journal = {Nat. Mater.},
  volume = {17},
  pages = {21--28},
  year = {2018},
  doi = {10.1038/nmat5031}

}

@article{Clark2018,
  author = {Clark, O. J. and Neat, M. J. and Okawa, K. and Bawden, L. and Markovi{'c}, I. and Mazzola, F. and Feng, J. and Sunko, V. and Riley, J. M. and Meevasana, W. and Fujii, J. and Vobornik, I. and Kim, T. K. and Hoesch, M. and Sasagawa, T. and Wahl, P. and Bahramy, M. S. and King, P. D. C.},
  title = {Fermiology and superconductivity of topological surface states in {PdTe$_2$}},
  journal = {Phys. Rev. Lett.},
  volume = {120},
  pages = {156401},
  year = {2018},
  doi = {10.1103/PhysRevLett.120.156401}
}

@article{Clark2019,
  author = {Clark, O J and Mazzola, F and Marković, I and Riley, J M and Feng, J and Yang, B-J and Sumida, K and Okuda, T and Fujii, J and Vobornik, I and Kim, T K and Okawa, K and Sasagawa, T and Bahramy, M S and King, P D C},
  title = {A general route to form topologically-protected surface and bulk Dirac fermions along high-symmetry lines},
  journal = {Electron. Struct.},
  volume = {1},
  pages = {014002},
  year = {2019},
  doi = {10.1088/2516-1075/ab09b7}
}

@article{Salis2021,
  author = {Salis, M. V. and Huang, Y. K. and de Visser, A.},
  title = {Heat capacity of type-I superconductivity in the Dirac semimetal {PdTe$_2$}},
  journal = {Phys. Rev. B},
  volume = {103},
  pages = {104502},
  year = {2021},
  doi = {10.1103/PhysRevB.103.104502}
 }

@article{McFarane2026,
  author = {McFarlane, Emily C. and Sanna, Antonio and Gilbert, Matthew J. and Krieger, Jonas A. and Date, Mihir and Domaine, Gabriele and Pal, Banabir and Chakraborty, Anirban and Sivakumar, Pranava K. and Constantinou, Procopios C. and Hartl, Anna and Della Valle, Enrico G. and Pellegrini, Camilla and Strocov, Vladimir N. and Parkin, Stuart S. P. and Schr{"o}ter, Niels B. M.},
  title = {Van Hove singularities, superconductivity, and the Josephson diode effect in {NiTe$_2$} and {PdTe$_2$}},
  journal = {Phys. Rev. Lett.},
  volume = {136},
  pages = {086401},
  year = {2026},
  doi = {10.1103/hp1t-zd7y}
}

@article{Pattalwar1988,
  author = {Pattalwar, S. M. and Dixit, R. N. and Shete, S. Y. and Basu, B. K.},
  title = {Low-temperature specific heat of {PdPb$_2$}},
  journal = {Phys. Rev. B},
  volume = {38},
  pages = {7067},
  year = {1988},
  doi = {10.1103/PhysRevB.38.7067}
}

@article{Arushi2021,
  author = {Arushi and Motla, K. and Kataria, A. and Sharma, S. and Beare, J. and Pula, M. and Nugent, M. and Luke, G. M. and Singh, R. P.},
  title = {Type-I superconductivity in single-crystal {Pb$_2$Pd}},
  journal = {Phys. Rev. B},
  volume = {103},
  pages = {184506},
  year = {2021},
  doi = {10.1103/PhysRevB.103.184506}
}

@article{Sharma2022,
  author = {Sharma, M. M. and Karn, N. K. and Rani, Poonam and Bhowmik, R. N. and Awana, V. P. S.},
  title = {Bulk superconductivity and non-trivial band topology analysis of {Pb$_2$Pd}},
  journal = {Supercond. Sci. Technol.},
  volume = {35},
  pages = {084010},
  year = {2022},
  doi = {10.1088/1361-6668/ac7c42}

}

@article{Mitra2017,
  author = {Mitra, S. and Okawa, K. and Kunniniyil Sudheesh, S. and Sasagawa, T. and Zhu, Jian-Xin and Chia, Elbert E. M.},
  title = {Probing the superconducting gap symmetry of {$\alpha$-PdBi$_2$}: A penetration depth study},
  journal = {Phys. Rev. B},
  volume = {95},
  pages = {134519},
  year = {2017},
  doi = {10.1103/PhysRevB.95.134519}
}

@book{Blaha2018,
  author = {Blaha, P. and Schwarz, K. and Madsen, G. K. H. and Kvasnicka, D. and Luitz, J. and Laskowski, R. and Tran, F. and Marks, L. D.},
  title = {{WIEN2k}: An Augmented Plane Wave Plus Local Orbitals Program for Calculating Crystal Properties},
  publisher = {Technische Universit{\"a}t Wien},
  year = {2018}
}

@article{Chen2001,
  author = {Chen, X. H. and Wang, Y. S. and Xue, Y. Y. and Meng, R. L. and Wang, Y. Q. and Chu, C. W.},
  title = {Correlation between the residual resistance ratio and magnetoresistance in {MgB$_2$}},
  journal = {Phys. Rev. B},
  volume = {65},
  pages = {024502},
  year = {2001},
  doi = {10.1103/PhysRevB.65.024502}
}

@article{Li2006,
  author = {Li, Qi and Liu, B. T. and Hu, Y. F. and Chen, J. and Gao, H. and Shan, L. and Wen, H. H. and Pogrebnyakov, A. V. and Redwing, J. M. and Xi, X. X.},
  title = {Large anisotropic normal-state magnetoresistance in clean {MgB$_2$} thin films},
  journal = {Phys. Rev. Lett.},
  volume = {96},
  pages = {167003},
  year = {2006},
  doi = {10.1103/PhysRevLett.96.167003}
}

@article{Okawa2018,
  author = {Okawa, K. and Kanou, M. and Namiki, H. and Sasagawa, T.},
  title = {Extremely large magnetoresistance induced by hidden three-dimensional Dirac bands in nonmagnetic semimetal {InBi}},
  journal = {Phys. Rev. Mater.},
  volume = {2},
  pages = {124201},
  year = {2018},
  doi = {10.1103/PhysRevMaterials.2.124201}
}

@article{Kohler1938,
  author = {Kohler, M.},
  title = {Zur magnetischen Widerstands{\"a}nderung reiner Metalle},
  journal = {Ann. Phys.},
  volume = {424},
  pages = {211--218},
  year = {1938},
  doi = {10.1002/andp.19384240124}
}

@article{Feng2025,
  author = {Feng, Y. and Wang, Y. and Rosenbaum, T. F. and Littlewood, P. B. and Chen, H.},
  title = {Fermi surface origin of the low-temperature magnetoresistance anomaly},
  journal = {Matter},
  volume = {8},
  pages = {102105},
  year = {2025},
  doi = {10.1016/j.matt.2025.102105}
}

@article{Xu2021Kohler,
  author = {Xu, Jing and Han, Fei and Wang, Ting-Ting and Thoutam, Laxman R. and Pate, Samuel E. and Li, Mingda and Zhang, Xufeng and Wang, Yong-Lei and Fotovat, Roxanna and Welp, Ulrich and Zhou, Xiuquan and Kwok, Wai-Kwong and Chung, Duck Young and Kanatzidis, Mercouri G. and Xiao, Zhi-Li},
  title = {Extended Kohler's rule of magnetoresistance},
  journal = {Phys. Rev. X},
  volume = {11},
  pages = {041029},
  year = {2021},
  doi = {10.1103/PhysRevX.11.041029}
}

@article{Pate2024,
  author = {Pate, Samuel and Chen, Bowen and Shen, Bing and Li, Kezhen and Zhou, Xiuquan and Chung, Duck Young and Divan, Ralu and Kanatzidis, Mercouri G. and Welp, Ulrich and Kwok, Wai-Kwong and Xiao, Zhi-Li},
  title = {Extended Kohler's rule of magnetoresistance in {TaCo$_2$Te$_2$}},
  journal = {Phys. Rev. B},
  volume = {109},
  pages = {035129},
  year = {2024},
  doi = {10.1103/PhysRevB.109.035129}
}

@article{Sonika2025,
  author = {Bagga, Sonika and Gangwar, Sunil and Kumar, Pankaj and Taraphder, Arghya and Yadav, C. S.},
  title = {Extended Kohler's scaling and isosbestic point in the charge density wave state of {1T-VSe$_2$}},
  journal = {J. Phys.: Condens. Matter},
  volume = {37},
  pages = {195601},
  year = {2025},
  doi = {10.1088/1361-648X/adc6e4}

}

@article{Chen2025Ni,
  author = {Chen, Bowen and Zhang, Hengyuan and Li, Jingyuan and Hu, Deyuan and Huo, Mengwu and Wang, Shuyang and Xi, Chuanying and Wang, Zhaosheng and Sun, Hualei and Wang, Meng and Shen, Bing},
  title = {Unveiling the multiband metallic nature of the normal state in the nickelate {La$_3$Ni$_2$O$_7$}},
  journal = {Phys. Rev. B},
  volume = {111},
  pages = {054519},
  year = {2025},
  doi = {10.1103/PhysRevB.111.054519}

}
\end{document}